# Dynamics from elastic neutron-scattering via direct measurement of the running time-integral of the van Hove distribution function

*Antonio Benedetto*[*,1,2,3,4,5] *and Gordon J. Kearley*[2]

[1]*School of Physics, University College Dublin, Dublin 4, Ireland*
[2]*School of Chemistry, University College Dublin, Dublin 4, Ireland*
[3]*Conway Institute of Biomolecular and Biomedical Research, University College Dublin, Dublin 4, Ireland*
[4]*Department of Sciences, University of Roma Tre, Rome, Italy*
[5]*Laboratory for Neutron Scattering, Paul Scherrer Institut, Villigen, Switzerland*

*Corresponding author: antonio.benedetto@ucd.ie

**Abstract:**
We present a new neutron-scattering approach to access the van Hove distribution function directly in the time domain, I(t), which reflects the system dynamics. Currently, I(t) is essentially determined from neutron energy-exchange. Our method consists of the straightforward measurement of the running time-integral of I(t), by computing the portion of scattered neutrons corresponding to species at rest within a time *t, (*conceptually elastic scattering). Previous attempts failed to recognise this connection. Starting from a theoretical standpoint, a practical realisation is assessed via numerical methods and an instrument simulation.

# 1. Introduction:

In this letter we present a new method for probing single-particle dynamics, which are central to the study of atomic and molecular motion across many areas of research. These dynamics can be fully described by the van Hove self-distribution function, $G_{Self}(r,t)$, which represents the probability that a species has diffused a distance *r* over time *t*. Equivalently, by its spatial Fourier transform, I(Q,t), which represents the probability that a species is within a volume $4/3\pi(2\pi/Q)^3$ after a time *t*, where $\hbar Q=2\pi\hbar/r$ is the momentum transfer[1]. For our discussion we ignore the Q-dependence of the function, but for generality we retain "van Hove" to denote I(t). Probing these dynamics at molecular or atomic scales requires the use of techniques such as X-ray and neutron scattering. Neutrons are unique in being scattered by the atomic nuclei, rather than measuring the response of the electron clouds to the nuclear motion, as is the case in most other spectroscopies. This simplifies the analysis, and allows scattering contrast to be varied by isotopic composition, especially hydrogen and deuterium that are particularly important in polymers, solvation, soft matter systems and biosystems.

Current neutron-scattering spectroscopies for measuring these dynamics rely on the exchanged energy of each scattered neutron as shown by Brockhouse[2]. By scanning the energy difference between the neutron beam incident on the sample and that scattered by the sample, the scan is made with respect to one of the energies, the other energy being fixed[3-5]. Neutron spin-echo is different, but still encodes exchanged energy, here via neutron spin, providing I(t)[6], whereas other methods only provide S(ΔE), the Fourier transform of I(t), where ΔE is the neutron-energy exchange.

Rather than these "inverse" approaches, we propose an approach that uses the proportion of neutrons scattered elastically within different observation times, $t_{obs}$ (i.e. different energy resolutions), that is the probability that particles are stationary (within the time resolution of the measurement) after time $t=t_{obs}$, to accesses I(t) directly. This is fundamentally different, and our contention is that direct access to the van Hove I(t) function should have significant advantages in instrument design, and data analysis, in particular for soft matter systems and biosystems of high (structural and dynamical) complexity.

We will show that although measurement of I(t) may be difficult, measurement of its running time-integral up to time $t=t_{obs}$ (that we will denote the van Hove integral vHI(t)), is surprisingly straightforward. Fig. 1 gives a schematic illustration of the concept, and provided that a suitable route to the derivative of the vHI(t) is used, I(t) is obtained without fitting, modeling or Fourier transform.

This letter first presents the underlying theory, and then supports this by a numerical simulation. Counting-errors are then included to approach the real experiment, and finally an assessment using a Monte-Carlo instrument simulation illustrates one possible realisation of an instrument. The supplementary materials (SM) provides more detail of the theoretical approach, the protocol to extract I(t) from the measured vHI(t), and the instrument design.

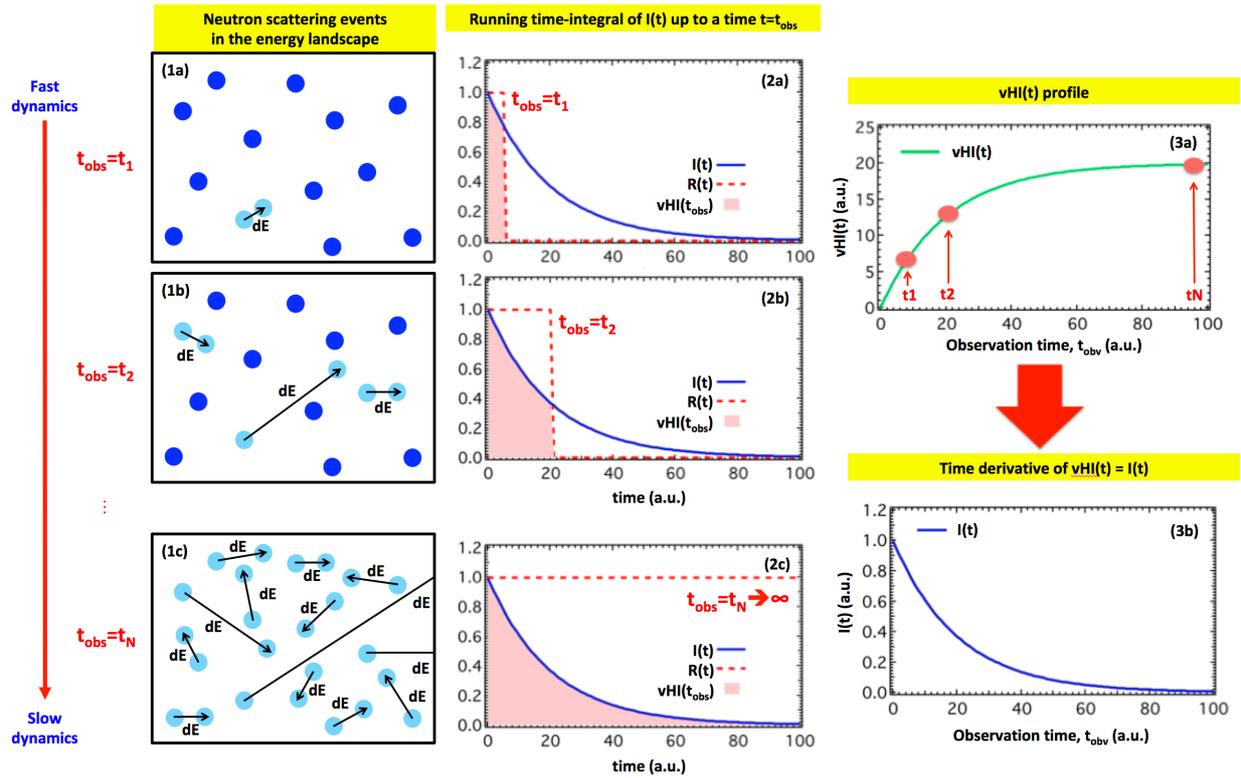

Fig. 1 – Illustration of the concept. Column 1 sketches the energy landscape of a system of particles, at three different observation times, $t_{obs}$. $t_{obs}$ is the time-resolution of the measurement and is inversely proportional to the instrumental energy-resolution. At short $t_{obs}$ only the rapid motions are detected, most of the system appearing at rest (1a). At intermediate $t_{obs}$ other motions are detected (1b), and for long $t_{obs}$ the slower motions are also detected (1c). Existing techniques require determination of many exchanged energy values, ΔE, to access I(t), typically operating at fixed $t_{obs}$. In general, our approach of obtaining the proportion of particles "at rest", as a function of $t_{obs}$, should be more efficient. This proportion formally corresponds to the running time-integral of I(t) of Eq. 2, that is the van Hove integral vHI(t=$t_{obs}$), as sketched in column 2, in which each $t_{obs}$ determines the upper integration limit of I(t). Differentiation of the measured vHI(t) (3a) provides I(t) directly (3b).

## 2. The state-of-the-art in "Elastic neutron scattering for dynamics":

Before introducing our theoretical framework, we will outline here related earlier work that tried to connect "elastic scattering" to I(t) and to system dynamics in general.

To the best of our knowledge the first attempt was made by Doster and co-workers in 2001[7-9]. These authors suggested the measured profile of elastic scattering vs energy resolution to be equal to I(t).

A second attempt was made in 2011[10,11] in which other authors show that an inflection point occurs in the (lin-log) plot of elastic neutron scattered intensity vs energy resolution at the point where the instrumental observation time matches the system overall relaxation time. This approach has been referred to as Resolution Elastic Neutron Scattering (RENS). It was clear that the RENS spectrum was not I(t), and consequently it was shown that the approach of Doster and co-workers (above mentioned) is conceptually incorrect.

Later in 2016[12] we proposed two new instrument concepts (suitable, respectively, for continuous and pulsed sources) designed with the specific purpose of measuring the RENS profile directly, i.e. elastic intensity vs energy resolution. This instrument would enable the overall relaxation time to be measured without Fourier transform or modeling. However, the instrument layouts, imposed a mismatch between the resolutions of the primary and secondary spectrometers. It turns out that having one of these resolutions much finer than the other is exactly what is required to access I(t) from the elastic intensity vs energy resolution. This forms the central part of this Letter, and we call this new spectroscopy for dynamics "Elastic Scattering Spectroscopy "(ESS) to distinguish from RENS and the other previous attempts.

Over the last decade, other experimental works used "elastic scattering" to access the system dynamics[13-15], also with the support of computations[16,17].

### 3. The new theoretical framework:

The starting point in the new theoretical approach we propose is to quantify "what is actually measured", i.e. the number of neutrons counted at the detectors, in these general types of neutron-scattering experiments. In doing so, instrumental features that contribute to the neutron-energy uncertainties are naturally included in the theory.

In this new framework, the number of elastically-scattered neutrons counted at a specific instrumental condition is a definite time-integral containing the van Hove I(t) function (see SM Paragraphs 1 and 2 for more detail):

$$N_{neutron}(Q, \Delta E = 0) = \int_0^\infty I(Q,t) R(t; \omega_R, \Delta\omega_R) F(t; \omega_R, \Delta\omega_F) dt \quad (1)$$

Equation 1 is quite general because it is the product of:
- the incident beam, R, the incoming distribution with average energy, $\hbar\omega_R$, and uncertainty, $\hbar\Delta\omega_R$;
- the sample, I, which is the van Hove function;
- the filter analyser, F, the probability that the average energy $\hbar\omega_R$ passes (where $\hbar\Delta\omega_F$ is the width of the distribution).

All functions are properly normalized (see SM Paragraph 1 for details).

The incident-beam function, R, convolutes with the exchange processes of the sample. Normally, neutron spectroscopies then scan this energy-exchange by varying $\omega_R$ either in R or in F, resulting in a double convolution. In our case however, we operate entirely in the elastic regime, that is keeping fixed $\omega_R$, and measure vHI(t) by changing either $\hbar\Delta\omega_R$ or $\hbar\Delta\omega_F$, and the second convolution is absent.

For a fixed instrumental condition, Eq. 1 is the "elastic scattering", which has been routinely measured as a function of system parameters such as temperature, pressure, hydration, etc[13,18-26]. In the present case however, we change the width of either the incident-beam, $\hbar\Delta\omega_R$, or the filter, $\hbar\Delta\omega_F$, by values that provide an incremental change in the observation time, $t_{obs}$, that is inversely proportional to the varied (and broader) energy-width, i.e. $t_{obs}=1.66/\Delta\omega$ (see SM Paragraph 2.3). When the fixed width is much less than the varied width, the measurement provides vHI(t) at incremental $t_{obs}$ times between $t_{min}$ and $t_{max}$, corresponding to the broadest and narrowest varied energy-widths, i.e. $t_{min}=1.66/\Delta\omega_{max}$ and $t_{max}=1.66/\Delta\omega_{min}$. It is conceptually easier to consider the case in which the incident-beam width is varied, and the analyser width is fixed (Eq. 2), although Eq. 1 shows the inverse approach to be equally valid:

$$N_{neutron}(Q,\Delta E=0) = \int_0^\infty I(Q,t)R(t;\omega_R,\Delta\omega_R)dt \approx \int_0^{t_{obs}=1.66/\Delta\omega_R} I(Q,t)dt \equiv vHI(t_{obs}) \qquad (2)$$

Ideally, vHI(t) is made using a step function for R (Eq. 2). The time integral of Eq. 2 then runs from $t=0$ to $t=t_{obs}$, giving vHI($t_{obs}$).

**4. Numerical validation:**

The practical R- and F-functions, however, are more likely to be Gaussian, and a crucial question that we address here is how this affects the applicability of Eq. 1. To do so, we start with a numerical simulation of the experiment in which we successively evaluate Eq. 3 (which is equivalent to Eq. 1, see SM) over a range of $t_{obs}$ to reproduce an input function from the time domain.

$$N_{neutron}(Q,\Delta E=0) = \frac{\int_{-\infty}^\infty [S(Q,\Delta E) \otimes R(\omega;\omega_R,\Delta\omega_R)] \cdot F(\omega;\omega_R,\Delta\omega_F)d\omega}{\int_{-\infty}^\infty S(Q,\Delta E)d(\Delta E) \cdot \int_{-\infty}^\infty F(\omega;\omega_R,\Delta\omega_F)d\omega} \qquad (3)$$

The first term of Eq. 3 is obtained from a chosen input I(t) function which is numerically Fourier transformed to the energy domain, S, and numerically convoluted with a (Gaussian) incident-beam function, R, which represents a primary monochromation device. The resulting function is multiplied with a much narrower Gaussian function, F, which represents a band-pass filter centred around the zero energy-transfer. All functions were normalised. The resulting integral, N(Q,ΔE=0), is stored as the value of the vHI(t=$t_{obs}$). vHI(t) is obtained via a step-wise change of $t_{obs}$. Because we can achieve this without counting errors, we can differentiate vHI(t) numerically and compare this result with the initial I(t) input function.

The comparison has been made for three representative input functions: (i) single exponential decay, (ii) sums of exponentials, and a (iii) stretched exponential. These describe: (i) a simple isotropic translational diffusion motion, (ii) a combination of two distinct isotropic relaxations, and (iii) a continuous distribution of relaxation processes (e.g. as in complex systems like proteins), respectively. Because any dynamical relaxation, including the one described by a stretched exponential, can be expanded in a sum of single relaxations, our chosen three cases offer an almost complete picture of any potential real case. Figure 2 shows that the agreement between the input and computed functions is good overall for all these three cases. The differences originate by having as the R-function a Gaussian function rather that a step function. We would contest that for investigations on real systems, the errors entailed by the non-ideal R- and F-functions would not significantly affect the conclusions. Moreover, if required, these errors could be estimated or possibly calibrated out (e.g. by numerically computing Eq. (2) for the specific case).

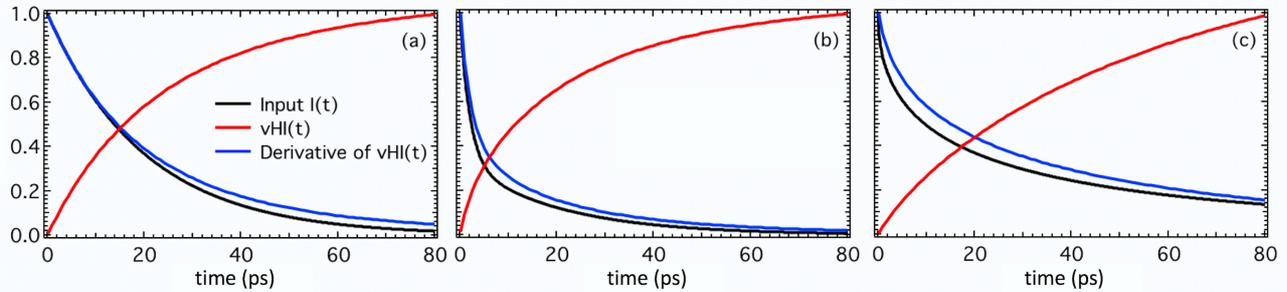

Fig. 2 – Numerical validations. The three plots are: (a) single exponential (with tau = 20 ps), (b) double exponential (with $tau_1$ = 20 ps, and $tau_2$ is twice as intense with = 2 ps), and (c) stretched exponential (with tau = 20 ps, and beta = 0.6). The other relevant parameters are: 20 <ℏΔ$ω_R$< 15000 micro-eV corresponding to 0.127 <$t_{obs}$< 83 picoseconds; ℏΔ$ω_F$ = 10 micro-eV. There are no counting errors, which allows the consequences of using a Gaussian for R and F to be assessed.

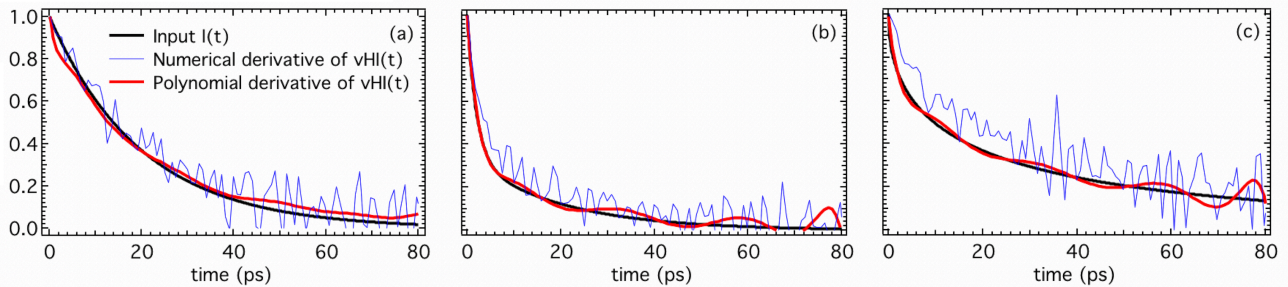

Fig. 3 – Numerical validation with counting error (described in the text). The three plots are: (a) single exponential, (b) double exponential, and (c) stretched exponential. The numerical derivative of vHI(t) is intractable, but its polynomial derivative reproduces the input function well. Ranges are as in Fig. 2.

To approach a real experiment, we have introduced counting errors on vHI(t). Based on the count-rate for a single point in a fixed-window scan on existing spectrometers[13] we estimate that an integral of $10^6$ counts for the whole energy spectrum, S, would correspond to about 3 minutes per vHI($t_{obs}$)-point. The transmission of the final-energy filter is only ~5%. Fig. 3 shows the results for the same three representative input functions of Fig. 2, illustrating that where counting errors are significant, the numerical derivative of the vHI(t) can become intractable. The problem of obtaining the derivative of noisy data is well known, and a number of solutions is available. We will consider the special case of polynomials, and then in the SM a simple Gaussian error-reduction method. Here, the polynomial approach not only provides a convenient route to the derivative, but is also physically meaningful.

Figure 3 shows that the agreement between the input I(t) function and the derivative of the vHI(t) using the polynomial method is good overall, but clearly has some fluctuations. In order to assess these, we have taken the numerical cosine Fourier transform of the energy spectrum S, with the equivalent counting statistics. Errors were set for each point in the energy spectrum, with the integral of this spectrum being $10^6$ counts as above. Fig. 4 shows that for most of the range the error propagation is similar in the two methods, and is probably the best that can be achieved. Similarly, we have used a simple Monte Carlo approach to reduce Gaussian errors during the numerical derivative (Fig. 4), which again gives approximately the same agreement with the input function. We conclude that there is a number of approaches to obtaining the derivative from the "experimental" vHI(t) that reduce the consequences of counting-errors to those of standard methods that require the whole energy spectrum.

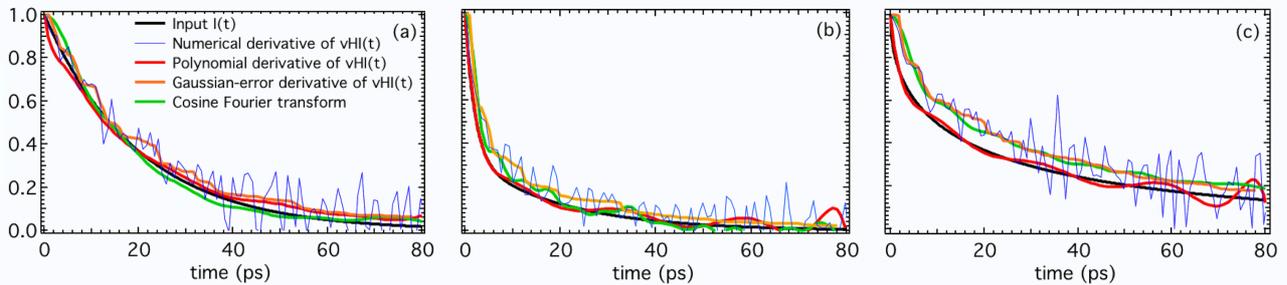

Fig. 4 – Numerical validation with counting error of several approaches to obtain I(t) from the "experimental" vHI(t). The three plots are: (a) single exponential, (b) double exponential, and (c) stretched exponential. The numerical derivative of vHI(t) is intractable, but its polynomial derivative and Gaussian-error derivative reproduce the input function well. The cosine FT is also shown for a consistency check. This figure is supposed for extending the results shown in Fig. 3 by showing that several approaches to get I(t) from vHI(t) are possible.

## 5. "Experimental" validation:

We can make one final step towards a practical instrument by estimating effects of beam inhomogeneities, placement of instrument components, and some account of the scattering processes. Several designs are feasible, but readers are referred to Ref. [12] for details of our recent McStas[27] instrumental design that we will use in this discussion. This consists of a standard backscattering geometry spectrometer with a stationary focussing monochromator. The distance from the monochromator to the sample is varied, and the focussing is adjusted to maintain the sample at the focal point. This causes the variation of $\Delta\omega_R$ (so $t_{obs}$) at the sample. A crystal filter of fixed width $\Delta\omega_F \ll \Delta\omega_R$ selects the narrow range of neutron-counts to be integrated (SM Fig. 5). Two samples were run. Firstly, vanadium which is a purely incoherent scatterer with no measurable processes on the time-scale of interest. This was used to determine the effect of the finite energy-width of the filter, $\Delta\omega_F$, and other instrumental-errors. Secondly, a sample with a scattering process equivalent to a single exponential decay with tau = 200 ps. Twenty-five incrementally spaced observation-times, $t_{obs}$, were selected by changing the focal-length of the monochromator. The ratio of the elastic intensity to the total intensity in the scattered beam was determined to obtain vHI(t=$t_{obs}$). Data-treatment consisted of taking the numerical derivative of the sample vHI(t) and dividing by that of the vanadium. Note that in this case the numerical derivatives were used directly. The results in Fig. 5 illustrate an exponential-decay function that fits the data, in reasonable agreement with the input. For more details please refer to "*SM Paragraph 5*" first, and then to Ref. [12] directly.

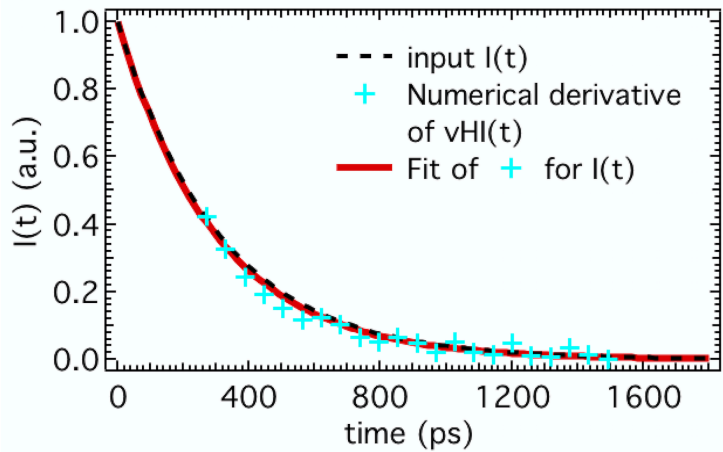

Fig. 5 – "Experimental" validation by a McStas simulation (of a new instrument designed to access vHI(t) directly). The 25 points (cyan) are the "measured" I(t) obtained by the numerical derivative of the "measured" vHI(t). Each of those 25 points has been measured (i) at 25 different monochromator-to-sample distances corresponding to 25 different value of $\Delta\omega_R$ (so $t_{obs}$), (ii) but with fixed value of $\Delta\omega_F$, as per Eq. (1). The best fit of the "experimental" I(t)-points (red curve) agrees very well with the input I(t) (dashed black curve).

## 6. Conclusions and remarks for the future:

In summary, the van Hove distribution function, I(t), can be used to describe the dynamics of physical systems, and traditionally neutron scattering has accessed this using energy-exchange measurements. However, we show that I(t) can be obtained directly from the proportion of species that have remained static within time $t$ (conceptually elastic scattering). More specifically, we have shown that, if a mismatch between the primary and secondary spectrometers is maintained, the elastic neutron-scattering intensity as a function of observation time (i.e. instrumental energy resolution) corresponds to the running-time integral of I(t) (which we denote as van Hove integral vHI(t)). The measured vHI(t) can easily be compared directly with the increasingly common molecular-dynamics simulations, or used as its derivative, I(Q,t).

This new theoretical result has been successfully tested by a series of numerical simulations, and an *in silico* experiment carried out on McStas on an *ad hoc* instrumental design. To distinguish our new method from previous attempts, we refer to it as "Elastic Scattering Spectroscopy, ESS".

Overall, the energy-exchange and vHI methods are equivalent, but in many instances one will have practical advantages over the other. To date the focus has been entirely on the energy-transfer method so it is likely that there are instances where our vHI method would be better.

At the moment, none of the available neutron spectrometers worldwide can be used to measure vHI(t) and then access I(t) as proposed here. Recently, we proposed *ad hoc* instrument concepts to do so, and we hope these will motivate better new designs that will be built. We believe that our new ESS approach can impact the use of neutron scattering for dynamics in the several cases of study in which either the complexity of the systems and/or their low availability and/or the sample environment requirements make standard approaches for dynamics, as QENS and NSE, difficult. This could be certainly the case for soft matter and bio- systems. Finally, we also believe that extremely simple instrumental designs can be achieved, leading/opening the way to compact versions of the instrument optimized to work for small neutron sources based on radio-frequency quadrupole accelerators (RFQ).

**Acknowledgment:**

The Authors acknowledge support from the University College Dublin (UCD) under the Seed Funding Scheme, with additional support provided by UCD Schools of Chemistry and Physics. A.B. acknowledges support from Science Foundation Ireland (grant no. 15-SIRG-3538) and the Italian Ministry of Education, University and Research (grant no. MIUR-DM080518-372).

**Additional information:**

**Correspondence** to antonio.benedetto@ucd.ie

*Supplementary information* accompanying this paper are at the bottom, after the bibliography.

*Author contributions:* Antonio Benedetto and Gordon J. Kearley contributed equally on all the major aspects of this work. Antonio Benedetto and Gordon J. Kearley wrote the manuscript text, and prepared the figures.

*Competing interests:* The Authors declare no competing interests in relation to this work.

# Supplementary Information for "Dynamics from elastic neutron-scattering via direct measurement of the running time-integral of the van Hove distribution function"

*Antonio Benedetto*[*,1,2,3,4,5] *and Gordon J. Kearley*[2]

[1]*School of Physics, University College Dublin, Dublin 4, Ireland*
[2]*School of Chemistry, University College Dublin, Dublin 4, Ireland*
[3]*Conway Institute of Biomolecular and Biomedical Research, University College Dublin, Dublin 4, Ireland*
[4]*Department of Sciences, University of Roma Tre, Rome, Italy*
[5]*Laboratory for Neutron Scattering, Paul Scherrer Institut, Villigen, Switzerland*

Corresponding author: antonio.benedetto@ucd.ie

## Supplementary information

This supplementary material (SM) is intended to complete and support the main letter by presenting in more details the theoretical framework, and the data processing protocol.

## 1. The most general theoretical framework:

Let us consider the most general layout of a neutron scattering experiment: neutrons come from the source via a neutron guide and are energy-shaped in the primary spectrometer. They then hit the sample and are scattered into the analyser of the secondary spectrometer before reaching the detector bank (SM Fig. 1a).

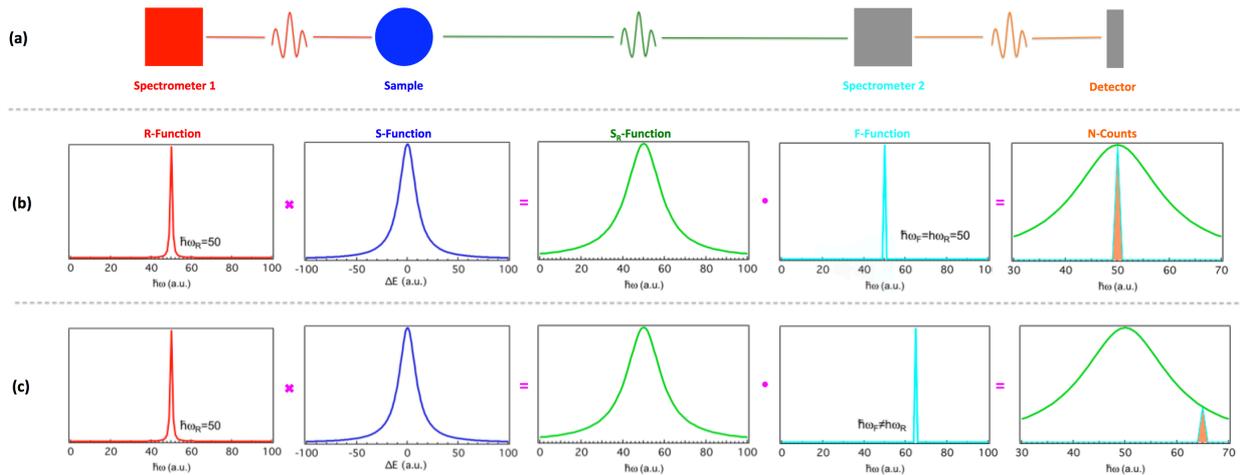

SM Fig. 1 – (a) The most general layout of a neutron scattering experiment in which the interplay between the primary and the secondary spectrometer enables the measurement of either the elastic (b) or the inelastic (c) contribution. Following Brockhouse[2], it is the inelastic scattering that has been routinely used to access system dynamics, this being done by changing either $\omega_R$ or $\omega_F$, to measure an energy exchange. We propose a new method to access the system dynamics by collecting the purely elastic contribution as a function of either $\Delta\omega_R$ or $\Delta\omega_F$.

Neutrons from the primary spectrometer can be mathematically described by the distribution function of their energies, $R(\omega;\omega_R,\Delta\omega_R)$, which is often a near symmetric function, centred around a specific energy value, $\hbar\omega_R$, with a variance, $\hbar\Delta\omega_R$. In many cases, this is a Gaussian function but others such as a Lorentzian or triangular function, also exist. The energy integral of R is the total number of neutrons coming from the primary spectrometer that hit the sample.

The sample can be mathematically described as the distribution function of the exchanged energies of the system (atomic and molecular) processes, $S(Q,\Delta E)$, or, by its time Fourier transform $I(Q,t)$ known as intermediate scattering function, or, by its space-time Fourier transform $G(r,t)$ known as the van Hove (total) distribution function. Ignoring the detailed balance, this near symmetric sample-component is centred at $\Delta E=0$, and has a distribution of inelastic events. If present, the symmetric broadening of the elastic component is called quasi-elastic scattering, and its width and intensity are related to the system dynamics. The sample function is a probability distribution that is normalised.

When a neutron with energy ℏω$_R$ hits the sample, it can be scattered by a nucleus of the system atoms, with which it may exchange energy and momentum. Assuming single scattering, the scattered neutron-energy, ℏω$_{NEW}$, is related to a single scattering event in which the energy ΔE$_{Exchanged}$ passes between the neutron and the system, such as ℏω$_{NEW}$=ℏω$_R$-ΔE$_{Exchanged}$. This energy exchanged corresponds to an allowed energy change of the system. In this respect, S(Q,ΔE) can be seen as the probability distribution that an incoming neutron will exchange an energy ΔE with the system, so changing its energy from ℏω$_R$ to ℏω$_{NEW}$=ℏω$_R$-ΔE$_{Exchanged}$.

After the scattering process, the energy distribution is described by the convolution of the energy distribution incident-beam, R(ω;ω$_R$,Δω$_R$), with the distribution of the system-exchanged energies, S(Q,ΔE):

$$S_R(Q,\omega;\omega_R,\Delta\omega_R) = S(Q,\Delta E) \otimes R(\omega;\omega_R,\Delta\omega_R) \qquad (1)$$

S$_R$(Q,ω;ω$_R$,Δω$_R$) is referred to as the "scattering law', which is a symmetric function as S(Q,ΔE), but now centred at ℏω$_R$ and with a different width taking into account the spread in the incoming neutron-energies, ℏΔω$_R$, which includes the uncertainty of the incident energy. For this reason, R is usually called the energy resolution-function of the (primary) spectrometer. The total area of S$_R$ integrated in energy and space is equal to the number of scattered neutrons. For a perfectly monochromatic incident-beam, R=δ(ω-ω$_R$), then S$_R$=S(Q,ℏω=ΔE+ℏω$_R$). Clearly, in this case the energy distribution after the scattering process maps the system exchanged energies.

The scattered beam then reaches the secondary spectrometer. Only the neutrons fulfilling the instantaneous time and space conditions of the secondary-spectrometer analysis are detected. For this single measurement (as opposed to a scan that we will discuss later) the secondary spectrometer acts as a filter. If F(ω;ω$_F$,Δω$_F$) describes the probability that a neutron with energy ℏω$_F$ passes through this filter, the number of neutrons reaching the detectors is:

$$N_{neutron} = \int_{-\infty}^{\infty} S_R(Q,\omega;\omega_R,\Delta\omega_R) F(\omega;\omega_F,\Delta\omega_F) d\omega \qquad (2)$$

Usually F is a symmetric function centred around a specific energy value, ℏω$_F$, and with a certain width, ℏΔω$_F$. For example this is a delta-function for a perfect crystal analyser, and only the neutrons with the exact matching energy will be detected. The F-function is a probability distribution that is normalised, i.e. its energy integral has to be equal to 1.

By combining Eq. 1 and Eq. 2, and taking into account the normalisations detailed above, the total number of neutron reaching the detector is:

$$N_{neutron} = \frac{\int_{-\infty}^{\infty} [S(Q,\Delta E) \otimes R(\omega;\omega_R,\Delta\omega_R)] \cdot F(\omega;\omega_F,\Delta\omega_F) d\omega}{\int_{-\infty}^{\infty} S(Q,\Delta E) d(\Delta E) \cdot \int_{-\infty}^{\infty} F(\omega;\omega_F,\Delta\omega_F) d\omega} \quad (3)$$

By using the convolution theorem, and assuming that the two probability distribution-functions are normalized and even, we have:

$$N_{neutron} = \frac{\int_{-\infty}^{\infty} [S(Q,\Delta E) \otimes R(\omega;\omega_R,\Delta\omega_R)] \cdot F(\omega;\omega_F,\Delta\omega_F) d\omega}{\int_{-\infty}^{\infty} S(Q,\Delta E) d(\Delta E) \cdot \int_{-\infty}^{\infty} F(\omega;\omega_F,\Delta\omega_F) d\omega} =$$

$$= \int_{-\infty}^{\infty} [S(Q,\Delta E) \otimes R(\omega;\omega_R,\Delta\omega_R)] \cdot F(\omega;\omega_F,\Delta\omega_F) d\omega =$$

$$= \int_{-\infty}^{\infty} \left[ \int_{0}^{\infty} I(Q,t) R(t;\omega_R,\Delta\omega_R) e^{-i\omega t} dt \right] \cdot F(\omega;\omega_F,\Delta\omega_F) d\omega =$$

$$= \int_{0}^{\infty} dt \int_{-\infty}^{\infty} d\omega I(Q,t) R(t;\omega_R,\Delta\omega_R) F(\omega;\omega_F,\Delta\omega_F) e^{-i\omega t} = \quad (4)$$

$$= \int_{0}^{\infty} I(Q,t) R(t;\omega_R,\Delta\omega_R) \left[ \int_{-\infty}^{\infty} F(\omega;\omega_F,\Delta\omega_F) e^{-i\omega t} d\omega \right] dt =$$

$$= \int_{0}^{\infty} I(Q,t) R(t;\omega_R,\Delta\omega_R) F(t;\omega_F,\Delta\omega_F) dt \quad (5)$$

Eq. 4 and its integral in energy, i.e. Eq. 5, describe the most general neutron scattering experiment. Eq. 5, in particular, shows the symmetric role played by the first and the second spectrometer for a single measurement. Basically, the number of neutrons reaching the detector banks is unchanged by interchanging the primary and the secondary spectrometers.

SM Fig. 1b and 1c show how Eq. 5 works for the elastic and inelastic scattering, respectively. To show the concept more clearly, the R- and F-functions have been approximated to delta functions. In the purely elastic regime, $\hbar\omega_R = \hbar\omega_F$, only elastically scattered neutrons arrive at the detector (SM Fig. 1b); whereas in the inelastic regime, only the neutrons with energy exchanges $\Delta E = \hbar\omega_R - \hbar\omega_F$ are detected. In the inelastic regime by varying either $\hbar\omega_R$ or $\hbar\omega_F$, it is possible to scan the whole $S_R$ (in green), and then indirectly access S (in blue) by de-convoluting with R or F (in red and cyan, respectively), and I(Q,t) by time-FT. In our method, however, we access the I(Q,t) directly in time domain by operating in the purely elastic regime without any need of FTs and de-convolutions.

In the following we will show in more detail how the standard methods operate and how they can be obtained from our general approach of Eq. 5, and we present our elastic method for dynamics.

## 2. Our method vs "standard" methods:

*2.1 "Standard" methods for dynamics:*

As pointed out at the end of the previous paragraph, in the inelastic regime it is possible to access $S_R$ by varying either $\hbar\omega_R$ or $\hbar\omega_F$. This is the method currently used in neutron scattering to access system dynamics, as proposed by Brockhouse for which he received the 1994 Nobel Prize in Physics[2]. Eq. 5 shows the symmetric role played by the primary and the secondary spectrometer for a single measurement. Thus, it is possible to use either a fixed monochromatic neutron-beam and scan the neutron final energies by varying $\hbar\omega_F$ (SM Fig. 2a), or use to vary the distribution of incoming neutrons by varying $\hbar\omega_R$ and detect a specific energy (SM Fig. 2b).

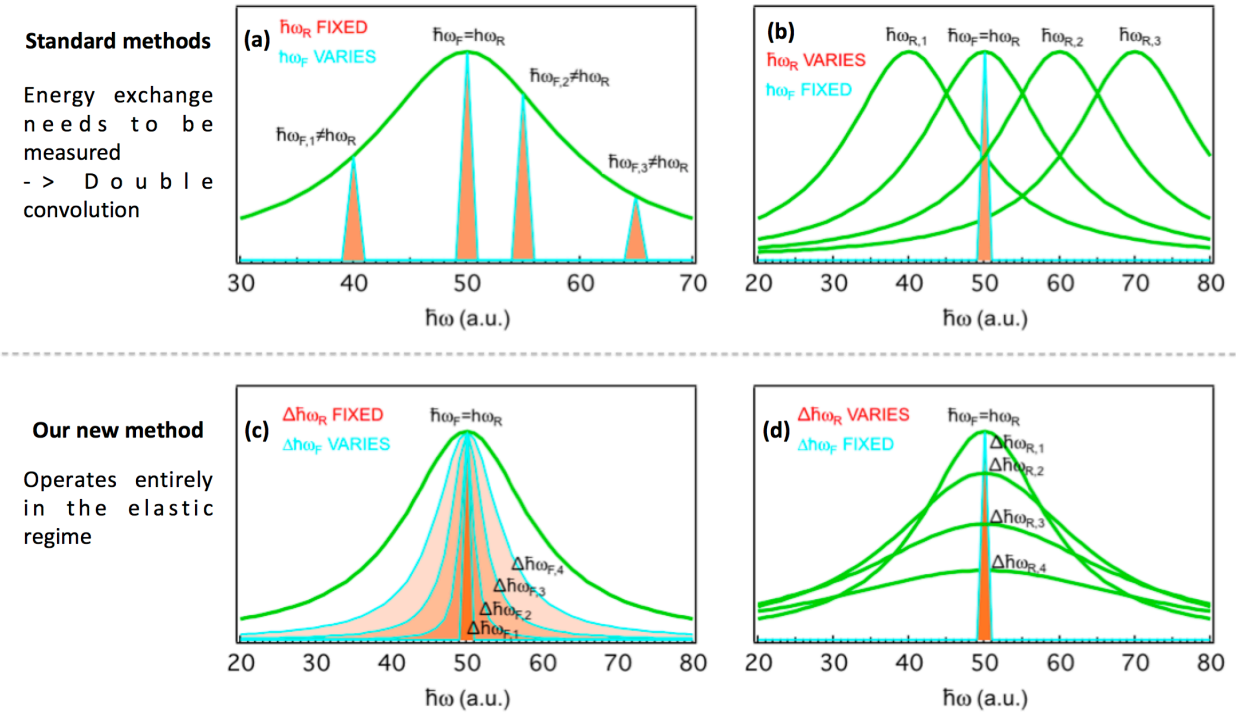

SM Fig. 2 – Comparison between "standard" methods for dynamics (a,b) and our new method (c,d). Standard methods operating in the inelastic regime, need to compute the energy exchange, and result in a double convolution to access S(Q,ΔE). The neutron energy-uncertainties, $\hbar\Delta\omega_R$ and $\hbar\Delta\omega_F$, are usually fixed. Our method is fundamentally different. It operates in the purely elastic regime ($\omega_R=\omega_F$), accessing the running time-integral of the I(t) directly by scanning either $\Delta\omega_R$ or $\Delta\omega_F$. Eq. 5 shows the symmetric role of the two spectrometers: In the standard methods when the energy of the primary spectrometer is fixed, then by varying the energy of the secondary spectrometer it is possible to scan S(Q,ΔE) (a), or when it is the energy of the secondary spectrometer to be fixed, then S(Q,ΔE) can be scanned by changing the energy of the first spectrometer (b). Here the double convolution is apparent. In contrast, our method relaxes one of the energy uncertainties, either $\hbar\Delta\omega_F$ (c) or $\hbar\Delta\omega_R$ (d), with the other being fixed to a narrower value. This accesses the portion of elastic scattering which corresponds to the running-time integral of I(t) as per Eq. 8 and 12, respectively. Case (d) is the one we considered in the main text.

These two protocols are basically equivalent. However, it is easier initially to focusing first on the former, summarised in SM Fig. 2a. The neutrons at the detector have energy-exchange of $\Delta E=\hbar\omega_R-\hbar\omega_F$. By varying $\hbar\omega_F$ and keeping $\hbar\omega_R$ fixed, it is possible to measure the number of neutrons as a function $\Delta E$ (see the four peaks in SM Fig. 2a). This number is by definition $S_R$. Normally the energy uncertainties of the primary and secondary spectrometers are kept similar, i.e. $\hbar\Delta\omega_R \approx \hbar\Delta\omega_F$. From a mathematical point of view, this protocol results in a convolution between $S_R$ and F. Considering that $S_R$ is itself a convolution, S×R, it results in a double convolution (as it is well known). It is easier to see the second convolution in the second symmetric protocol of SM Fig. 2b. In this case, $\hbar\omega_F$ is fixed, and the $S_R$ is scanned by varying $\hbar\omega_R$. This results in energy shifts of $S_R$ being centred at the different $\hbar\omega_R$ (see the different $S_R$, in green, in SM Fig. 2b). In practice a standard backscattering spectrometer operates exactly in this way. A time modulation of $\hbar\omega_R$ is generated via a doppler device, and after the scattering process, a crystal analyser reflects only the neutrons with a certain energy $\hbar\omega_F$.

We highlight that these standard approaches measure $S_R$. Then the first step is to de-convolute it with R to get S. This de-convolution process is usually done numerically and by proposing a function that is convoluted with R to obtain $S_R$. Once a good approximation of S is computed this has to be time-FT to get I(Q,t). There are obvious advantages for methods able to access I(Q,t) without the need of either FTs or de-convolutions, accessing I(Q,t) directly in the time domain.

*2.2 Our method for dynamics:*

Our approach is based on the consideration that the best way to probe system dynamics is via I(Q,t) directly rather than $S_R$, and that Eq. 5 contains I(Q,t) already, so we consider the most direct access to I(Q,t) based on Eq. 5.

In our approach we suggest scanning the system dynamics via the species that are at rest within a certain time frame (elastic regime) rather the exchanged energies (inelastic regime) as summarised in Fig. 1 in the main text. Thus, in our proposed set-up $\hbar\omega_R=\hbar\omega_F$, either $\hbar\Delta\omega_R$ or $\hbar\Delta\omega_F$ varies, and Eq. 5 gives:

$$N_{neutron}(Q,\Delta E=0;\Delta\omega_R,\Delta\omega_F) = \int_0^\infty I(Q,t)R(t;\omega_R,\Delta\omega_R)F(t;\omega_R,\Delta\omega_F)dt \qquad (6)$$

Eq. 6 is the core of our approach, and shows the two symmetric and equivalent ways in which it can be implemented.

One way (given in the main text and summarized in SM Fig. 2d) uses a very narrow filter, i.e. $\hbar\Delta\omega_F<<\hbar\Delta\omega_R$. This means that the F-function in the time domain can be approximated to unity in the time interval in which the R-function is different form zero, giving:

$$N_{neutron}(Q,\Delta E=0;\Delta\omega_R) = \int_0^\infty I(Q,t)R(t;\omega_R,\Delta\omega_R)dt \qquad (7)$$

Eq. 7 shows that the time integral of I(Q,t) can be scanned via Δω$_R$ with the primary spectrometer. In the ideal case of a step R-function, it simply corresponds to the definite running-time integral of the van Hove I(Q,t) function up to t$_R$, i.e. the van Hove integral function vHI(t=t$_R$):

$$N_{neutron}(Q,\Delta E = 0; t_R) = \int_0^\infty I(Q,t) H(t_R - t) dt = \int_0^{t_R} I(Q,t) dt \equiv vHI(t_R) \tag{8}$$

where t$_R$ is defined as resolution time and is proportional to the inverse of Δω$_R$. More precisely it is possible to show that (see SM paragraph 2.3):

$$t_R = \frac{1.66}{\Delta \omega_R} \tag{9}$$

As a result, a plot of the derivate of the number of counts against t$_R$ is the van Hove I(Q,t) function:

$$\frac{d}{dt_R}\{N_{neutron}(Q,\Delta E = 0; t_R)\} = \frac{d}{dt_R}\left\{\int_0^{t_R} I(Q,t) dt\right\} = I(Q,t) \tag{10}$$

In the main text we have shown numerically and also by an *in-silico* experiment that Eq. 8 holds reasonably well in real cases where the R-function is not a step function. We will also show later in this SM that the condition ℏΔω$_F$<<ℏΔω$_R$ allows a partial relaxation of Δω$_F$ (SM Fig. 3), which has practical implications in improving counting statistics.

In the converse arrangement (summarized in SM Fig. 2c) we use highly monochromated incident beam corresponding to a very narrow R-function. In the condition of ℏΔω$_R$<<ℏΔω$_F$, the R-function in the time domain can be approximated to unity during the time interval in which the F-function is non-zero, giving:

$$N_{neutron}(Q,\Delta E = 0; \Delta \omega_F) = \int_0^\infty I(Q,t) F(t; \omega_R, \Delta \omega_F) dt \tag{11}$$

Eq. 11 has the same form of Eq. 7, but we are now scanning the filter width to measure the van Hove integral. Also in this case, with the ideal case of a step function as F-function, it simply corresponds to the definite running-time integral of the van Hove I(Q,t) function up to t$_F$, i.e. the van Hove integral function vHI(t=t$_F$):

$$N_{neutron}(Q,\Delta E = 0; t_F) = \int_0^\infty I(Q,t) H(t_F - t) dt = \int_0^{t_F} I(Q,t) dt \equiv vHI(t_F) \tag{12}$$

where $t_F$ is defined as filter time and is proportional to the inverse of $\Delta\omega_F$ as $t_R$ in Eq. 9, i.e. $t_F=1.66/\Delta\omega_F$. Also in this case, a plot of the derivate of the number of counts against $t_F$ is the van Hove $I(Q,t)$ function:

$$\frac{d}{dt_F}\{N_{neutron}(Q,\Delta E=0;t_F)\} = \frac{d}{dt_F}\left\{\int_0^{t_F} I(Q,t)\,dt\right\} = I(Q,t) \tag{13}$$

In summary, our method is based on the direct measurement of the van Hove integral, vHI(t), by measuring the portion of the system that is moving more slowly than the sensitivity of the experimental set-up by scanning the time range with either $\Delta\omega_R$ or $\Delta\omega_F$. Then the van Hove function $I(Q,t)$ can be extracted by computing the derivative of this integral as described in the main text and in this SM paragraph.

*2.3 From the instrumental energy resolution, Δω, to the (instrumental resolution) observation time, $t_R$:*
In this paragraph the relationship between the "instrumental resolution time, $t_R$", and the "instrumental energy resolution, $\Delta\omega_R$", anticipated in Eq. 9 is derived. The resolution time has been also referred to as "observation time, $t_{obs}$". We are deriving this relationship for the most common case of Gaussian function as R-function. As mention above, R has to be normalized in the ω-space to the total number of incoming neutrons, N, (Eq. 14):

$$R(\omega,\Delta\omega=1.177\sigma) = \frac{N}{\sqrt{2\pi}\sigma}Exp\left[-\frac{\omega^2}{2\sigma^2}\right] = \frac{N}{\sqrt{2\pi}\frac{\Delta\omega}{1.177}}Exp\left[-\frac{\omega^2}{2\left(\frac{\Delta\omega}{1.177}\right)^2}\right] \tag{14}$$

$$\int_{-\infty}^{\infty} R(\omega,\Delta\omega)\,d\omega = N \tag{15}$$

with σ as the variance of the distribution and $\Delta\omega=1.177\sigma$, as commonly used in experiments, is the HWHM of the distribution. The time-FT of Eq. 14 gives R in the time domain (Eq. 16):

$$R(t;\Delta\omega) = \frac{1}{\sqrt{2\pi}}\int_{-\infty}^{\infty} R(\omega;\Delta\omega)Exp[-i\omega t]\,d\omega = \frac{N\frac{\Delta\omega}{1.177}}{\sqrt{2\pi}}Exp\left[-\frac{t^2\left(\frac{\Delta\omega}{1.177}\right)^2}{2}\right] \tag{16}$$

By re-arranging the argument of the exponential, Eq. 9 linking the energy resolution to the observation time can be eventually determined (Eq. 18):

$$R(t;t_R=1.66/\Delta\omega) = \frac{N\frac{\Delta\omega}{1.177}}{\sqrt{2\pi}}Exp\left[-\left(\frac{t}{\frac{1.66}{\Delta\omega}}\right)^2\right] = \frac{N\frac{\Delta\omega}{1.177}}{\sqrt{2\pi}}Exp\left[-\left(\frac{t}{t_R}\right)^2\right] \tag{17}$$

$$t_R = \frac{1.66}{\Delta\omega} \tag{18}$$

It is useful at this point to remember the transformation factor form *energy* to *time*:

$$\frac{\Delta\omega[\mu eV]}{\Delta\omega[ps^{-1}]} = 657.9 \; \mu eV \cdot ps \qquad (19)$$

To conclude this paragraph, the same approach as followed above can be used to adapt Eq. 9 to different real cases of R-functions. Moreover, a numerical evaluation of Eq. 8 with the actual resolution function of the particular spectrometer can be used to estimate or possibly calibrate out the discrepancy originated by not using a step function for R (see main text).

*2.4 Neutron Spin Echo:*

We want to conclude this theoretical framework by showing that Eq. 4 can also be used for determining the neutron spin-echo (NSE) case. NSE is currently used to access system dynamics. However, even though it accesses the I(Q,t) directly, it entails the energy exchanged (via the neutron spin). For this reason we have included NSE in the "standard" approaches.

In NSE the primary energy-resolution is very narrow, this basically means that the R-function can be considered equal to unity in the time interval in which the F-function is different form zero. Moreover, filtering in NSE is done with a polariser filter, giving for the F-function a cosine function, $\cos(\omega \cdot t_F)$, where $t_F$ is known as Fourier time and depends on instrumental parameters. Finally, NSE is usually used in the echo-condition, which in our framework means $\omega_R = \omega_F$. These inserted into Eq. 4 give back the basic NSE equation, i.e. Eq. 20:

$$N_{neutron} = \frac{\int_0^\infty dt \int_{-\infty}^\infty d\omega I(Q,t) R(t;\omega_R,\Delta\omega_R) F(\omega;\omega_F,\Delta\omega_F) e^{-i\omega t}}{\int_{-\infty}^\infty S(Q,\Delta E) d(\Delta E) \cdot \int_{-\infty}^\infty F(\omega;\omega_F,\Delta\omega_F) d\omega} =$$

$$= \frac{\int_0^\infty dt \int_{-\infty}^\infty d\omega I(Q,t) \cos(\omega \cdot t_F) e^{-i\omega t}}{\int_{-\infty}^\infty S(Q,\Delta E) d(\Delta E)} =$$

$$= \frac{\int_{-\infty}^\infty d\omega \left[ \int_0^\infty I(Q,t) e^{-i\omega t} dt \right] \cos(\omega \cdot t_F)}{\int_{-\infty}^\infty S(Q,\Delta E) d(\Delta E)} =$$

$$= \frac{\int_{-\infty}^\infty S(Q,\Delta E = \hbar\omega) \cos(\omega \cdot t_F) d\omega}{\int_{-\infty}^\infty S(Q,\Delta E) d(\Delta E)} =$$

$$= \langle \cos(\omega \cdot t_F) \rangle = I(Q,t_F) \qquad (20)$$

In NSE $t_F$ is scanned by changing the magnetic field or the wavelength of the incoming neutrons.

*2.5 Final remarks on the theoretical framework:*

We conclude that by taking into account both the primary spectrometer R-function and the secondary spectrometer F-function, our theoretical framework is the most general (to the best of our knowledge). It allows us to introduce our method and distinguish it from the standard QENS and NSE approaches. Nevertheless both QENS and NSE can be determined formally in our general framework summarised in Eq. 5.

## 3. Matching primary resolution-width with analyser width:

In order to show the closest agreement between the input and calculated functions in the numerical simulation the analyser was chosen to have a fixed narrow band-pass, $\Delta\omega_F$, as suggested by the theory. In practice, if a larger difference can either be tolerated or calibrated out, then the width of the band-pass can be increased as some function of the incident energy width, $\Delta\omega_R$. In the limit, the two functions would be matched, and as discussed in the text this provides a final function that is rather different from the input, as illustrated in SM Fig. 3. We show two other cases in the figure where the incident-energy width and the band-pass of the filter decrease together, but with the band-pass width being held at 0.5 and 0.33 of the incident energy-width. In reality, the time required for measurements at longer system-times dominate the total counting time and the possible gains made by relaxing the filter for measurements at lower times are not as great as may have been expected.

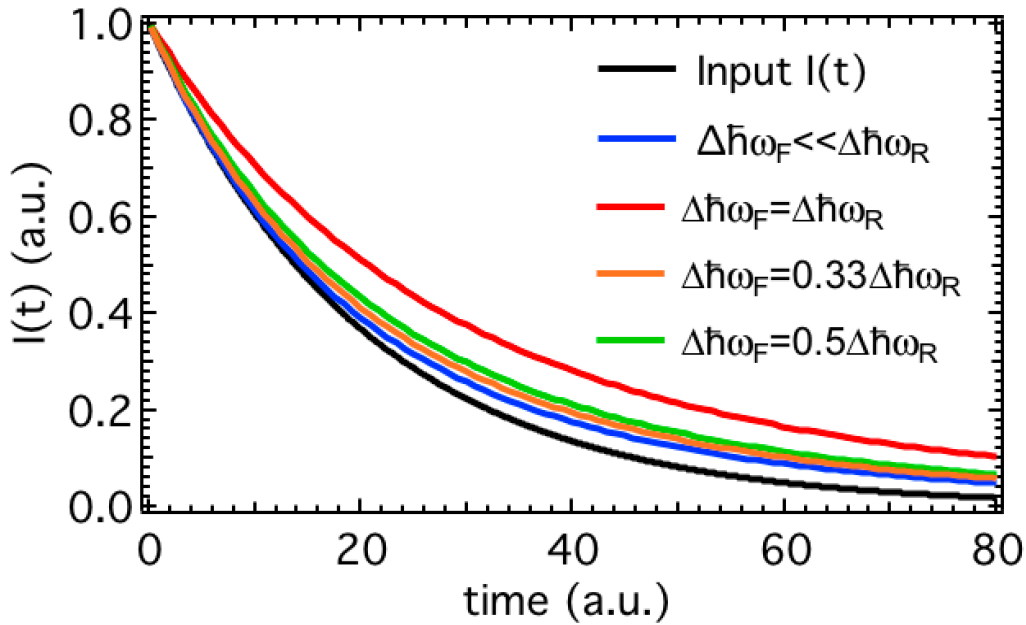

SM Fig. 3 – Effect of changing the energy width of the secondary spectrometer to match the energy width of the primary spectrometer.

## 4. Data analysis:

In this paragraph more details on the data analysis are provided. SM Fig. 4 summarises the numerical protocol used.

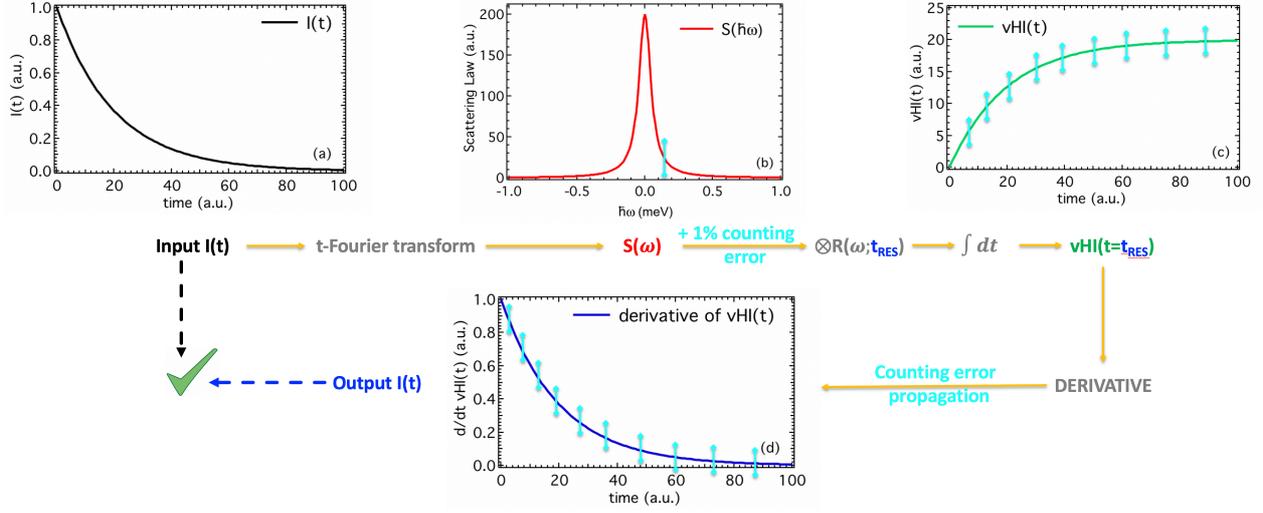

SM Fig. 4 – Summary cartoon of the numerical protocol. The numerical protocol is based on the numerical evaluation of Eq. 3 (main text) over a range of $t_{RES}$ to reproduce an input function from the time domain.

*4.1 Polynomial:*

In the main text we opted for a polynomial expansion to map the van Hove integral vHI(t) and then extract I(t). Our choice not only provides a convenient route to the derivative, but is also physically meaningful. This is because I(t) can generally be considered as some distribution of exponential functions, for which Taylor expansion is a real vector of the space of the power of time, $\{t^n\}$. As a result, if the vector ($a_0$, $a_1$, $a_2$, $a_3$, ..., $a_N$) represents the experimental profile, then the vector ($a_1$, $2a_2$, $3a_3$, ..., $Na_N$, 0) is its derivative, i.e. I(t):

$$vHI(t) = \sum_{i=0}^{N} a_i t^i \rightarrow I(t) = \sum_{i=1}^{N} i a_i t^{i-1} \tag{21}$$

By using a high-order polynomial to represent vHI(t), the coefficients for the polynomial of I(t) are obtained using Eq. 21 (Fig. 3 in the main text).

*4.2 Reduction of Gaussian errors:*

When the numerical derivative of experimental data with statistical errors is taken, the scatter on the result is much larger. To overcome such difficulty, we mapped the measured function in the vector space of the power of *time* by using the polynomial expansion described in the previous paragraph. Several other methods have been developed by the scientific community to overcome such problems as for example spline. In this paragraph, we present a simple numerical approach we developed to show how these errors can be reduced whilst taking the derivative. The basic assumption is that the errors on the data are normally distributed so that there must exist a vector of Gaussian-errors that when added to the data removes the scatter. By using a Monte-Carlo approach we try to reduce the scatter on the derivative by adding randomly chosen vectors of Gaussian errors to the experimental function. The cost-function to be minimised has 2 components. Firstly, the sum of the absolute values of the numerically-determined second derivative. The second cost stems from the requirement that the experimental errors are normally distributed, which also requires that width of the scatter between the raw experimental data and the simulated data should be $\sqrt{2\pi}$ multiplied by the height. This cost component is then the difference between the actual width and $\sqrt{2\pi}$. An empirical factor was used to balance the relative weight of these two components. In the main text, Fig. 4 reproduces Fig. 3 together with the derivative of the measured vHI obtained by the Gaussian error approach above-presented.

*4.3 Cosine FT of energy spectrum:*

The energy spectrum in the numerical simulation was cosine Fourier transformed using the Eq. 20 given for the spin-echo method as a simple consistency check that takes account of the truncation of the energy spectrum.

In the absence of counting errors this produces a smooth curve that agrees well with the input function, so we conclude that the range of our simulated data is adequate for validation purposes (not shown). Counting errors were then introduced on the energy spectrum in a way that is consistent with the errors on the vHI-function and the cosine Fourier transform of this noisy energy spectrum was then made (Fig. 4 in the main text). This results in oscillations in the time-domain function that are broadly comparable with both the function from the polynomial derivative, and the numerical noise-reduction method.

**5. Our McStas instrumental designs do measure vHI(t) directly:**

In Ref. [12] we have proposed two new instrument designs with the specific purpose to access the elastic neutron scattering intensity as a function of the energy resolution (of the primary spectrometer) directly. One design is suitable for continuous sources and is based on a standard backscattering geometry; the other is suitable for pulsed sources and is based on a time-of-flight option/approach. The former design, which have been used in this letter to "experimentally" validate our new theory, is reported for convenience in SM Fig. 5

This design is very similar to a standard backscattering spectrometer but with a focusing monochromator (as primary spectrometer) able to vary the energy resolution of the neutrons directed into the sample. This is achieved by changing the curvature of the monochromator and adjust accordingly its distance to the sample to maintain the focus. At the longest distance (position A, SM Fig. 5) the monochromator is almost flat, thus a perfect backscattering geometry is restored, and the best energy resolution is achieved (usually order of 1 micro-eV). By increasing the curvature of the monochromator, and decreasing accordingly its distance to the sample (maintaining the focus) poorer energy resolutions can be achieved, usually down to 100 micro-eV (positions B and C, SM Fig. 5). For instance, the time window of the accessible observation times is largely dictated by the longest and shortest distances available. As a result, the energy resolution is scanned by the first spectrometer, with the other parts of the instrument being identical to standard backscattering spectrometers: that is, the neutrons scattered by the sample do reach the analyzers', and only the "elastic" neutrons will be reflected to the detectors and counted.

The analyzer crystals are matched with the monochromator crystals: the perfect matching is at the fully backscattering condition which is at the longest monochromator-to-sample distance (position A, SM Fig. 5). For all the other distances, the matching is not perfect. This mis-match was perceived as a significant practical disadvantage in Ref. [12], however we now understand that it is vital for Eq. (6) to access I(t). Conventional matching of these widths precludes access to I(t), as reported in SM Fig. 3. As a result, both designs proposed in Ref. [12] can measure vHI(t), but only at the condition that $\Delta\omega_F \ll \Delta\omega_R$.

For the "experimental" validation presented in this Letter, vHI(t) at twenty-five incrementally spaced observation-times, $t_{obs}$, were collected by changing the monochromator-to-sample distance from 30 cm corresponding to $t_{obs}$= 350 ps to 270 cm corresponding to $t_{obs}$= 1600 ps, At each step the curvature of the monochromator was changed to maintain the focus on the sample, and the filter width was fixed at $\hbar\Delta\omega_F$= 1 micro-eV. The system relaxation time was 200 ps.

Finally, it is important to note the following. The McStas Monte-Carlo simulation of Ref. [12] was originally developed to determine the system overall relaxation time (in complex (bio-) systems) as proposed by the RENS protocol[10,11]. This time corresponds to the time of the inflection point in the elastic scattering intensity versus energy resolution profile: i.e. "there will be an inflection point in the lin-log plot of the elastic intensity versus the instrumental energy-resolution when the energy resolution matches the overall system relaxation-time"[10,11]. As a result, the aim of Ref. [12] was originally to develop new *ad hoc* instrumental designs to measure solely and directly the elastic intensity component of the scattering as a function of the instrumental energy resolution. Generally speaking, this "elastic intensity vs energy resolution" is not vHI(t), and that was not out aim in Ref. [12]. However, thanks to the new theory proposed here, it is now clear that at the condition that $\Delta\omega_F \ll \Delta\omega_R$, the measured "elastic intensity vs energy resolution" corresponds vHI(t). As a result, the two instrument designs of Ref. [12] can do much more than what was realised when they were designed: they can access the entirely I(t) directly in the time domain by measuring vHI(t). However, this can be achieved only at the condition that $\Delta\omega_F \ll \Delta\omega_R$, as dictated by Eq. (6). In light of this, the mis-match of the primary resolution and secondary filter-width, is a vital ingredient in the new elastic scattering spectroscopy we are proposing.

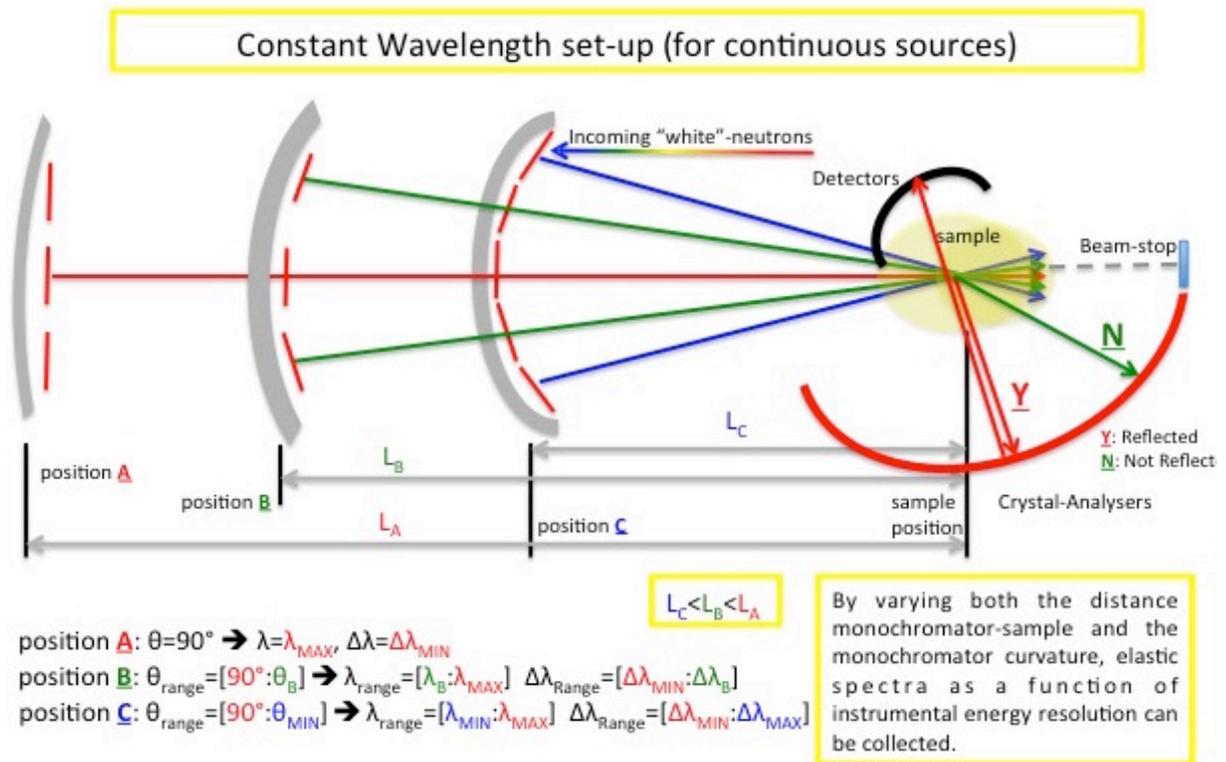

SM Fig. 5 – Lay-out of the spectrometer used in the McStas computation for the "experimental" validation of our new method. The different colours represent different wavelengths, which also have different energy resolutions, different velocities and, in turn, different TOF. Adapted from Ref. [12] and reproduced with the permission from the publisher.